\documentclass[prc,aps,nofootinbib,showkeys,showpacs,twocolumn,preprintnumbers]{revtex4-1}

\usepackage[T1]{fontenc} 
\usepackage{epsfig} 
\usepackage{graphicx} 
\usepackage{amssymb}
\usepackage{amsmath}
\usepackage[english]{babel}
\usepackage{color} 
\definecolor{darkgreen}{rgb}{0.2,0.5, 0.2}
\usepackage{ulem}
\usepackage{dcolumn,booktabs}
\newcolumntype{d}[1]{D{.}{.}{#1}}

\usepackage{lipsum}

\usepackage{dsfont}

\usepackage{hyperref}

\begin{document}

\preprint{NT@UW-23-02}

\title{Generation, dynamics, and correlations of the fission fragments' angular momenta}

\author{Guillaume Scamps$^{1}$}
\email{gscamps@uw.edu} 
\address{$^{1}$Department of Physics, University of Washington, Seattle, Washington 98195-1560, USA}
\author{George Bertsch$^{2}$}
\address{$^{2}$Institute for Nuclear Theory and Department of Physics, University of Washington, Seattle, Washington 98195-1560, USA}

\begin{abstract} 

The generation of angular momentum  in fissioning nuclei is not well
understood.  The predictions of different models
disagree, particularly 
concerning the correlation between the fragments' angular momenta.
In this article, a time-dependent collective Hamiltonian model is
proposed to treat the generation of the angular momentum in the fission fragments due to the 
quantum uncertainty principle as well as the dynamics of the collective wave function 
during and after scission. The model is constructed in the framework of the frozen Hartree-Fock approximation using a Skyrme energy
functional to extract deformations of the fission fragments as well as the
interactions in a derived collective Hamiltonian.  The fission
reactions studied are  $^{240}$Pu
$\rightarrow$ $^{132}$Sn+$^{108}$Ru and $^{240}$Pu $\rightarrow$
$^{144}$Ba+$^{96}$Sr.  
  The model can account for a large part of the
angular momentum found in experimental data.  
 The orientation of the angular momentum of each fragment is found to be mainly in the plane
perpendicular to the fission axis, in agreement with the experiment.  The
magnitudes of the
angular momenta in the two fragments are nearly uncorrelated, in
agreement with the recent experimental data of Wilson et al., Nature (London) 590, 566 (2021). 
Some of the conclusions of the traditional collective vibration model are supported by
the present model but some are not.
 Surprisingly, it is found that the angular momenta of the
fragments are slightly correlated positively as in a wriggling mode.  It is
also found that the presence of an octupole deformation in a fragment  can significantly increase the
generated angular momentum.
\end{abstract}

\maketitle


\section{Introduction}

The generation of angular momentum from fission fragments is a
complex phenomenon~\cite{Ben20}. 
Different
models have been introduced to calculate contributions to angular momentum including
statistical fluctuations~\cite{Mor80,Dos85,Ran21,Ran22} and
the quantum effects visible at scission \cite{Mik99,Bon07,Shn02,Ras69,Zie74}. 
 Recently, the projection method has been used to determine the intrinsic angular
distribution in the fission fragments
\cite{Ber19b,Mar21,Bul21,Bul22,Sch22}.  However, these methods
do not include quantum fluctuations or correlations between the
collective degrees of freedom (see, for example, Ref. 
\cite{Neg82,RS}). 

 The models can be separated into two kinds, the ones starting from nucleon degrees of freedom and those starting from collective degrees of freedom. In this article, we explore the dynamic effects in a collective quantum model.
Much of the angular momentum generated in fission can be understood with the
uncertainty principle.  When the fragments are deformed and aligned with the
fission axis, the radial orientation exhibits a peak at 0$^{\circ}$ with a
small fluctuation, $\Delta \theta$, and that requires a coherent superposition
of angular momentum, $L$.
This polarization is caused by a confining potential before scission. 
During scission, this confining potential disappears
progressively.  If scission occurs quickly, angular momentum distribution will remain unchanged.  However, if scission
occurs slowly, the evolution 
will be  adiabatic.
Therefore, it is essential
to incorporate a realistic treatment of the dynamics  to
calculate the evolution of the angular momentum during scission
and the Coulomb excitation phase.  Indeed, during and  after  scission,
the axes of deformation may be tilted with respect to the fission axis. 
They are then subjected to the quantum equivalent of the classical
torque due to the Coulomb\footnote{See the torque formulation in Appendix \ref{sec:torque}.}
interaction~\cite{Hof64,Wil72,Ras69,Mic99,Ber19a,Sca22}.

The present article proposes a model to describe the
time-dependent evolution based on a realistic interaction acting on the wave function of the orientation angles.  The goal is to
understand the fluctuations in the angular properties of the
post-fission fragment.  This approach is similar to the
Density-Constrained Hartree-Fock (Bogoliubov)
\cite{Cus85,Uma85,Sca19,God22} or Frozen Hartree-Fock (FHF)
approximations~\cite{Sim17,Uma21}, which have been used to model
accurately fusion tunneling.  
The model does not include the angular momentum
associated with quasiparticle excitations and so assumes that the fragments
are cold and rigid.  However, it goes beyond previous microscopic approaches 
in that it treats explicitly the rotational collective
degrees of freedom as well as their correlations.


The article is organized as follows. In Sec.  \ref{sec:method} we
present the collective Hamiltonian model and its ingredients.  In Sec. 
\ref{sec:generation}, the distributions of fragment angular
momenta are presented.  The correlations between the angular momenta of
the fragments are discussed in Sec.  \ref{sec:correlation}.  Sec. 
\ref{sec:MOI} discusses the sensitivity to parameters in the model. Sec. 
\ref{sec:octupole} discusses the contribution of octupole deformation
for the fission fragment $^{144}$Ba.

\section{Method}
\label{sec:method}

\subsection{Collective coordinates of the fissioning systems}

\begin{figure}[!h]
\centering
\includegraphics[width=.99\linewidth, keepaspectratio]{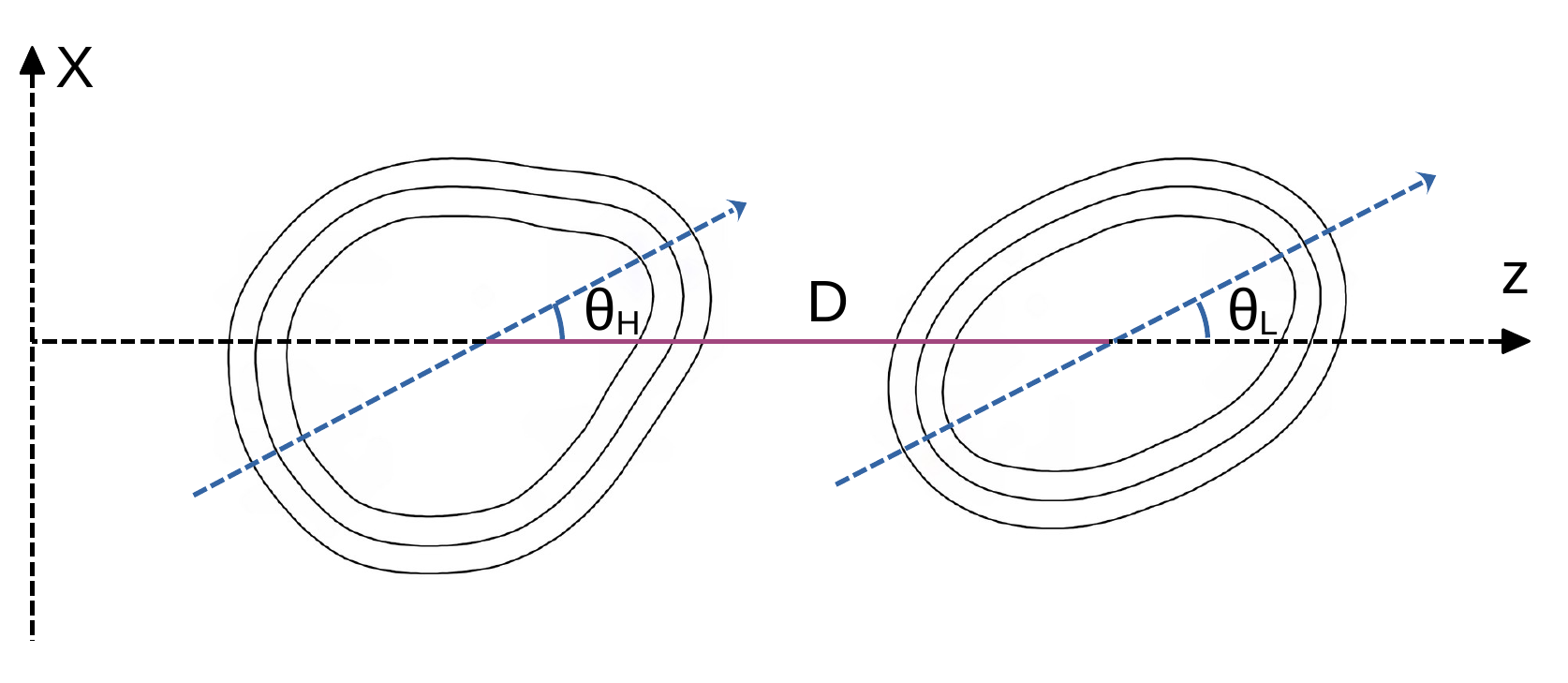}
\caption{  Schematic view of the model in the x-z plan in the case of $\varphi_H$=0 and  $\varphi_L$=0. The blue dashed line shows the principal axis of deformation of each fragment.} 
\label{fig:grah_2D_xyz}
\end{figure}

The model consists of two cold fragments separated by a distance $D$ along
the z-axis, which is treated as the fission axis.  The fragments
are oriented with their principal deformation axes forming an angle
$\theta$ with the z-axis and an azimuthal angle $\varphi$ with respect to
the x-axis as shown in Fig.  \ref{fig:grah_2D_xyz}.

To construct the collective Hamiltonian, the FHF method with the
Sly4d interaction~\cite{kim1997} described in a previous study
(Ref.~\cite{Sca22}) will be employed to calculate the interaction potential between the
fragments as a function of $D$ and the angular variables.
We treat two fission reactions: $^{240}$Pu
$\rightarrow$ $^{132}$Sn+$^{108}$Ru and $^{240}$Pu $\rightarrow$
$^{144}$Ba+$^{96}$Sr. The two mass splittings produce fragments
of quite different shapes.
The deformation is large for $^{108}$Ru, moderate for $^{96}$Sr,
and vanishes for the doubly magic $^{132}$Sn.  
The $^{144}$Ba has a large octupole deformation that turns out to have a significate 
role.  Due to its spherical shape, and the assumption that fragments are cold,
 the experimental spin of the $^{132}$Sn cannot be described in our model. 

\begin{table}[!h]
\caption{ Deformations of the four nuclei considered in this study,
calculated from the self-consistent ground state or superdeformed
minimum as described in the text.  The last column gives the
rigid moment of inertia on an axis perpendicular to the deformation axis of the nucleus. All fragments are found to be axially symmetric.}
\begin{tabular}{c|c|c|c|c|c}
\hline
\hline
 Nuc. & $\beta_2$ & $\beta_3$ & $\beta_4$& $E^*$ [MeV] & $I_{\rm Rigid}$ [$\hbar^2$/MeV] \\ %
\hline
$^{132}$Sn & 0. & 0. & 0. & 0 & 50.0\\
$^{108}$Ru & 0.82 & 0. & 0.71 & 3.5 & 51.4\\
$^{144}$Ba & 0.22 & 0.16 & 0.15 & 0 & 63.1 \\
$^{96}$Sr & 0.53 & 0. & 0.25 & 0 & 37.1 \\
\hline
\end{tabular}
\label{tab:def}
\end{table}

Except for $^{108}$Ru, the fragment deformations in the model are obtained from their  calculated
density distributions as isolated nuclei in their ground states. 
$^{108}$Ru is likely to be produced in a super-deformed minimum, and we
make that assumption.   The $\beta$ values are shown in  Table
\ref{tab:def}, using the definition\footnote{Note that the present definition of 
$Q_{\ell 0}$ differs by a factor 2 from the conventional formula for $\ell \neq 2$}
\begin{align}
\beta_{\ell} &= \frac{4 \pi }{3 A (r_0 A^{1/3})^{ \ell } }  Q_{l 0} \sqrt{\frac{2\ell+1}{16\pi}} ,\\
Q_{\ell 0}&= \int d^3 { \mathbf{r}} \,\rho(\mathbf{r})  r^{\ell} \sqrt{\frac{16\pi}{2 \ell +1}}  Y_{\ell,0}(\theta,\varphi) , \label{eq:mom}
\end{align}
with $r_0$~=~1.2~fm.  
%
The Hamiltonian also depends on the fragments' moments of inertia. 
Most of the calculations described below use the rigid values 
shown in
Table \ref{tab:def}, computed as
\begin{align}
I_{\rm Rigid} =  m \int d^3 {\mathbf{r}}  \,\rho(\mathbf{r}) (x^2 + y^2). 
\end{align}

The quantum degrees of freedom in the model are the orientation angles
of the fragments,
while the separation coordinate
between the fragments is treated as an external time-dependent parameter.  For
a full treatment of the angular momentum in the system, one would also
include the orientation of
the fission axis as another degree of freedom.  We assume that the system
as a whole has angular momentum zero in which case the fission axis
rotation is not an independent degree of freedom (see App. \ref{sec:ang_fis_axis}).

If one of the fragments
is spherical,  rotational invariance permits a Hamiltonian with only
one internal angular momentum variable, namely the angle between the 
fission axis and the axis of the deformed fragment.  When
both fragments are deformed, the Hamiltonians depend on three angular
variables, namely the two deformation axes 
$\theta_L$ and $\theta_H$ and the azimuthal angle
$\varphi = \varphi_H - \varphi_L$ between them.

\subsection{Model Hamiltonian}
\subsubsection{Single-angle model}

As mentioned above, the $^{132}$Sn+$^{108}$Ru fission reaction can be
treated with a single-angle Hamiltonian.  We write this as
\begin{align}
\hat H(D) &= \frac{\hbar^2}{2} \left( \frac{1}{I}+\frac{1}{I_{\Lambda}(D)} \right) \left( \frac{1}{\sin( \theta)} \frac{\partial}{\partial \theta} \left( \sin(\theta) \frac{\partial}{\partial \theta} \right) \right) \nonumber \\ 
&+ V( \hat \theta, D)
 \label{eq:Ham1D}
\end{align}
where the interaction $V( \hat \theta, D)$ includes both nuclear and
Coulomb terms. The rotational energy has two terms; $I$ is the inertia of
the fragment given in Table I and $I_\Lambda = m \frac{A_H A_L}{A_H+ A_L}D^2$ is the inertia of the
fragments about the system's center of mass. 
The Hamiltonian can be equivalently written in the angular momentum basis,
\begin{align}
H_{L L'}(D) &=  \frac{\hbar^2}{2} \left( \frac{1}{I}+\frac{1}{I_{\Lambda}(D)} \right) 
L(L+1) \delta_{LL'} + V_{L L'}( D) \label{eq:Ham1D_L},
\end{align}
which is more convenient to calculate.
The interaction is transformed to the angular momentum basis by
\begin{align}
V_{L L'}( D) = \int_0^{\pi} \sin(\theta) V( \theta, D) 
P'_L( \cos(\theta) ) P'_{L'}( \cos(\theta) )d \theta, \label{eq:pot_1D}
\end{align}
using the Legendre polynomial basis $P'_L(\theta)$ normalized such that, $\int_0^{\pi} \sin(\theta) P'_L( \cos(\theta) )^2 d \theta = 1$.
The potential $V( \theta, D)$ is determined in the Frozen Hartree-Fock  approximation (See Appendix
\ref{sec:pot1D}). It takes into account the long-range Coulomb potential and the short-range nucleus-nucleus interaction. 
The Hamiltonian is computed in a basis truncated at $L_{\rm max}=40~\hbar$.
The integration on Eq. \eqref{eq:pot_1D} is done with the simple trapezoidal rule with $\Delta \theta=\pi/500$.

\subsubsection{Triple-angle model}

When both fragments are deformed, as in  the $^{240}$Pu
$\rightarrow$ $^{144}$Ba+$^{96}$Sr case, three angles need to be specified
for the fragment orientation as mentioned in Sec. IIA. These correspond to
three quantum numbers in the angular-momentum basis states $|L_H, m , L_L,
-m \rangle$,  namely $L_H$ and $L_L$
for the principal quantum numbers of the fragments and $\pm m$ for the
azimuthal quantum numbers.  The total azimuthal angular momentum is zero
in the starting wave function (by assumption) and remains unchanged during the 
Schr\"odiner evolution.
The basic model assumption is again to treat the fission fragments as
rigid rotors using the distance 
$D$ between the fragments as a time-dependent
parameter of the Hamiltonian.  The collective Hamiltonian
can then be written,
\begin{equation}
\begin{aligned}
\hat H(D) &= \frac{\hbar^2}{2 I_H} \hat{L}_H^2 + \frac{\hbar^2}{2 I_L} \hat{L}_L^2 + \frac{\hbar^2}{2 I_{\Lambda}(D)} \hat{\Lambda}^2 + \hat V(D), \label{eq:Hamilt}
\end{aligned}
\end{equation}
with 
$\hat{\vec{\Lambda}} = - \hat{\vec{L_H}} - \hat{\vec{L_L}}$ the orbital angular momentum
of the fragments about the center of mass. 
The parameters $I_H$ and $I_L$ are the moment of inertia (MOI) associated with the rotors and $I_\Lambda(D)$ is the MOI associated with the angular momentum about the center of mass. The nucleus-plus-Coulomb potential  $V(\hat \theta_H,\hat \theta_L,\hat \varphi, D)$ (See Appendix \ref{sec:pot3D}) is calculated by the FHF approximation as a function of orientation angles and then converted in the $| L_H,L_L,m \rangle$ basis defined as,
\begin{widetext}
\begin{align}
| L_H, m, L_L, -m \rangle = \int_0^{\pi} & d \theta_H \int_0^{\pi} d \theta_L \int_0^{2\pi} d \varphi_H \int_0^{2\pi} d \varphi_L \sin(\theta_L) \sin(\theta_H)  \nonumber \\
 &  P'_{L_H,m}(\cos(\theta_H)) P'_{L_L,-m}(\cos(\theta_L)) e^{i m ( \varphi_H  - \varphi_L) } | \theta_H, \varphi_H, \theta_L, \varphi_L \rangle.
\end{align}
with $P'_{l,m}(x)$ the associated Legendre polynomials with the appropriate normalisation.

 The operator $\hat{\Lambda}^2$ can be written in the $| L_H,L_L,m \rangle$ basis,
\begin{align}
(\Lambda^2)_{L_H,L_L,m,L_H',L_L',m'} &=  \delta_{L_H,L_H'} \delta_{L_L,L_L'} \left( \frac{}{} \delta_{m,m'}  \left( L_H(L_H+1) + L_L(L_L+1)  -  2m^2 \right)  \right. \nonumber \\ 
&+         \delta_{m,m'+1} \sqrt{ (L_H(L_H+1) - m(m-1)) ( L_L(L_L+1) - m(m-1) ) }  \nonumber \\
&+ \left. \delta_{m,m'-1} \sqrt{ (L_H(L_H+1) - m(m+1)) ( L_L(L_L+1) - m(m+1) ) } \right).
\end{align}
\end{widetext}
Numerically, the space includes states with $L$ up to $30\hbar$ and $|m|\leq 2\hbar$, which gives a total number of states of 4443.  It has been checked that the results are unaffected by increasing the size of the basis.

\subsection{Schr\"odinger dynamics}
\label{sec:dynamics}

The system is evolved with the time-dependent Schr\"odinger equation,
\begin{align}
 i \hbar \frac{d}{dt} \Psi(t) = \hat H(D(t)) \Psi(t) \label{eq:TDSE}
\end{align}
taking $D(t)$ from 
time-dependent Hartree-Fock (TDHF) calculation in  Ref. 
\cite{Sca22}.  The starting point at $t=0$ is the ground eigenstate 
of the Hamiltonian at a separation distance $D_0$,  obtained by a direct diagonalization of the Hamiltonian in the angular momentum
basis. For most of the calculations
reported here we take $D_0 = 14 $ fm.  This was chosen
as  the closest point where the 
potential favors an aligned orientation of the
rotors along the fission axis.

 \begin{figure}[!h]
\centering
\includegraphics[width=.99\linewidth, keepaspectratio]{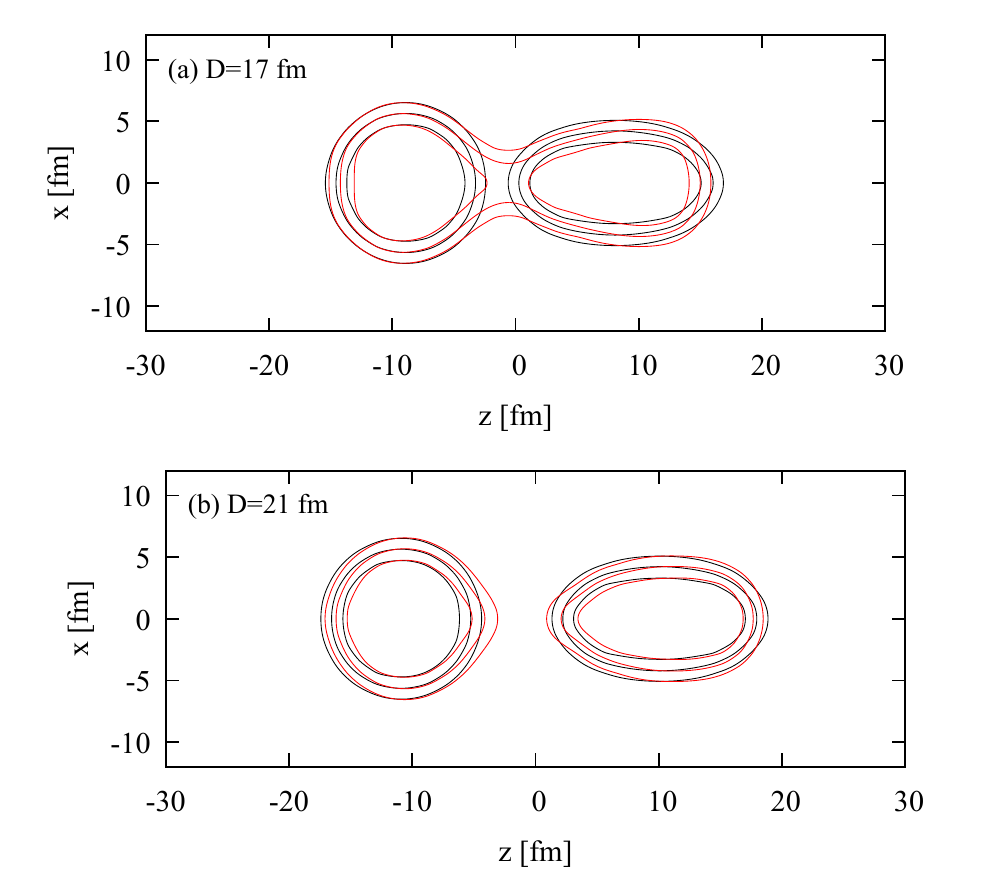}
\caption{ (Color online) Comparison between the static (black curves) and dynamical shapes obtained respectively by the FHF and TDHF calculations in the $^{240}$Pu $\rightarrow$ $^{132}$Sn+$^{108}$Ru case.
 } 
\label{fig:comp_shape_stat_dyn}
\end{figure} 
  
The evolution of
the wave packet $\Psi(t)$ is obtained by solving Eq.  \eqref{eq:TDSE} with
the Runge-Kutta method at the order 4 with a time step $\Delta t$ = 0.25
fm/c.  It has been checked that the final results were unaffected by a
change to a smaller value of  $\Delta t$.  The wave function is evolved
until $D(t)=100$ fm,  which is sufficient to have a complete convergence of
the final angular momentum distribution.

We found two problems with a naive use of the FHF nuclear interaction due
to the limited characterization of the nuclear shapes in the model.
The first one is just an annoyance  arising from the invariance of the fragment density under
the parity transformation when only even moments of the density are
present.  In that case, a pocket in the potential at $\theta=0$ will
be accompanied by an identical pocket at $\theta = \pi$.  This gives an
unwanted near-degeneracy of the eigenstates.  We deal with this by setting
the nuclear part of $V(D)$ beyond $\theta = \pi/2$ its value at $\pi/2$.

\begin{figure}[!h]
\centering
\includegraphics[width=.99\linewidth, keepaspectratio]{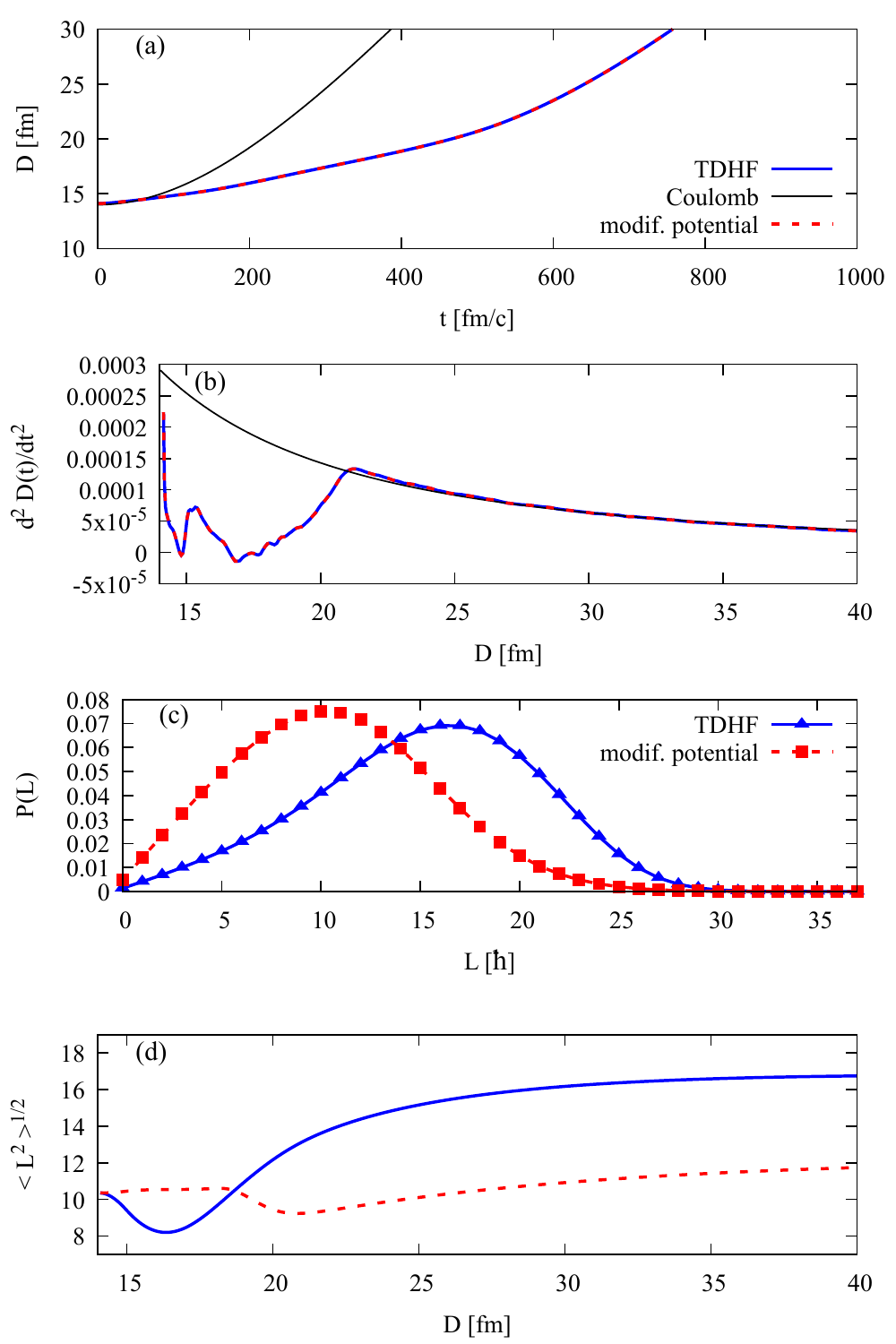}
\caption{ (Color online) Evolution of $D$ and angular momentum in the fission reaction $^{240}$Pu $\rightarrow$ $^{132}$Sn+$^{108}$Ru. 
a) Distance between the center of mass of the fragments as a function of time. b) Acceleration of the fragments. 
c) Final distribution of spin. d) Evolution of the average angular momentum as a function of the distance between the fragments.
 Different assumptions are made here: i) A TDHF evolution of $D(t)$ (blue curve), ii) a pure Coulomb repulsion from D=14 fm (black thin curve) iii) a TDHF evolution up to $D_{Sc}$=17 fm and pure Coulomb repulsion afterward iv) a TDHF evolution but the NN potential $V(\theta,D)$ is modified with $D$ $\rightarrow$ $D+4$ fm (red dashed curve). } 
\label{fig:effect_gamma_1D}
\end{figure} 

The second problem arises from the lack of explicit neck degrees of freedom
in the collective model.  During scission, the TDHF evolution proceeds through an elongation as well as a narrowing of the neck.  This gives rise to very high multipole
deformations in the newly formed fragments.  The FHF Hamiltonian
lacks those neck degrees of freedom, as  may be seen in Fig. \ref{fig:comp_shape_stat_dyn}
comparing the
density distributions at $D = 17$ fm.
The extended neck produces a stronger nuclear attraction between the 
nascent fragments than the FHF Hamiltonian provides.
We take this into account by displacing 
the FHF nuclear potential  in the
$D$ coordinate.  

The distance to shift the FHF potential is determined by matching the
final scission point as found in the TDHF calculation.  For the TDHF
calculation, we take that as the point in the dynamic evolution where
the acceleration of the fragments is maximal.  From Fig. \ref{fig:effect_gamma_1D}(b), this
occurs near $D =  21 $ fm.  For the FHF Hamiltonian, we take the
final scission point at the separation $D$ where the confinement
to small  orientation angles  disappears,  ie.

\begin{equation}
\frac{d^2 V_D(\theta)}{d \theta^2}\left.\right|_{\theta=0} = 0.
\end{equation} 

That point is at $D = 17$ fm,  4 fm closer.  This
is consistent with many other fission models
~\cite{Lem13,Lem19,Sta09,Sim14,Bul16},
locating the scission point  several fm beyond what one would calculate
in the frozen approximation.  The shift is incorporated in our
modified FHF $V'$ as follows:
\begin{align}
V'_{NN}(D)&= V_{NN,FHF}(D-4) \quad \text{for } D\geq D_0+4 \text{ fm}, \\
V'_{NN}(D)&= V_{NN,FHF}(D_0) \quad \text{for }D<D_0+4 \text{ fm}.
\end{align}
with $D_0 = 14$ fm.  We find that the modified interaction reduces the 
generation of angular
momentum (See Fig. \ref{fig:effect_gamma_1D}).  This is because stronger nuclear
interaction acts against the Coulomb interaction in the region 17 fm $<D<$ 21 fm.

\section{Calculated angular momentum of the fission fragments}
\label{sec:generation}
\subsection{Angular momentum in the one-angle model}

 \begin{figure}[!h]
\centering
\includegraphics[width=.99\linewidth, keepaspectratio]{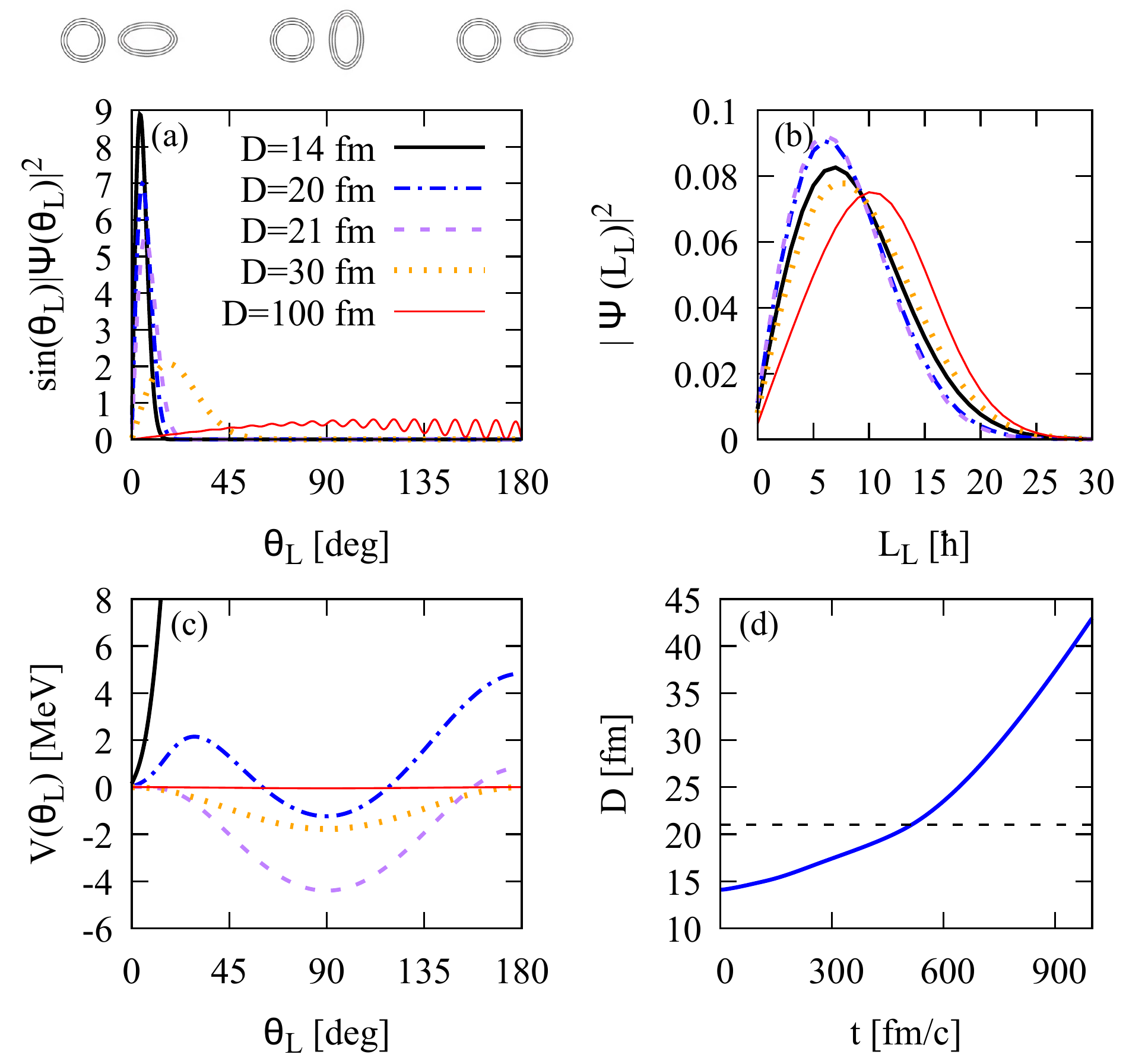}
\caption{ (Color online) (a) Angular wave function describing the orientation of the $^{108}$Ru light fragment at different distances $D$. The $^{132}$Sn complementary fragment is assumed to be spherical. (b) Snapshot of the angular momentum distribution at different times. (c) Angular potential in which the wave function is evolved. (d) Evolution $D(t)$ obtained from the TDHF calculation (blue solid curve), the dashed line shows the distance for which the acceleration of the fragment is the maximum. Above panel (a) a schematic figure shows the orientation of the light fragment as a function of the angle $\theta_L$.} 
\label{fig:Fig_expl_model_1D}
\end{figure}

In the case of the $^{240}$Pu $\rightarrow$ $^{132}$Sn+$^{108}$Ru fission, the $^{132}$Sn fragment is
spherical in the FHF approximation and $V(D)$ depends only on
$\theta_L$, the orientation angle of the
light fragment.  The initial probability distribution at
$D$=14 fm 
is shown in Fig.  \ref{fig:Fig_expl_model_1D} in the orientation
(panel (a)) and angular momentum basis (panel (b)).  The stiff
initial potential shown on panel (c) confines the wave function
in the small angle region around $\theta_L$=0.  The strongly
oriented wave function must include high angular momentum
components due to the uncertainty principle~\cite{Bon07,Fra04}. 
This initialization is similar to the model of ref. 
\cite{Bon07,Ras69,Zie74,Shn02,Mic99}.  However, the present model
is based on a realistic microscopic potential and does not rely on a small
angle approximation.

The evolution of the probability distribution in the one-angle
case is shown in Fig.  (\ref{fig:Fig_expl_model_1D}b).  Before
scission, the confining pocket at small angles becomes softer
and softer; the wave packet expands a bit, 
reducing the average angular momentum.  
After the scission at $D$=21 fm the wave packet spreads toward  $\theta_L$=90$^{\circ}$ due to the kinetic terms in the Hamiltonian as seen in Fig. (\ref{fig:Fig_expl_model_1D}a) and the Coulomb torque generates additional angular momentum (see Sec. \ref{sec:torque}).
The final angular momentum is obtained from the wave function at $D(t)=100$ fm.

 \subsection{Triple-angle calculation}

\begin{figure}[!h]
\centering
\includegraphics[width=.99\linewidth, keepaspectratio]{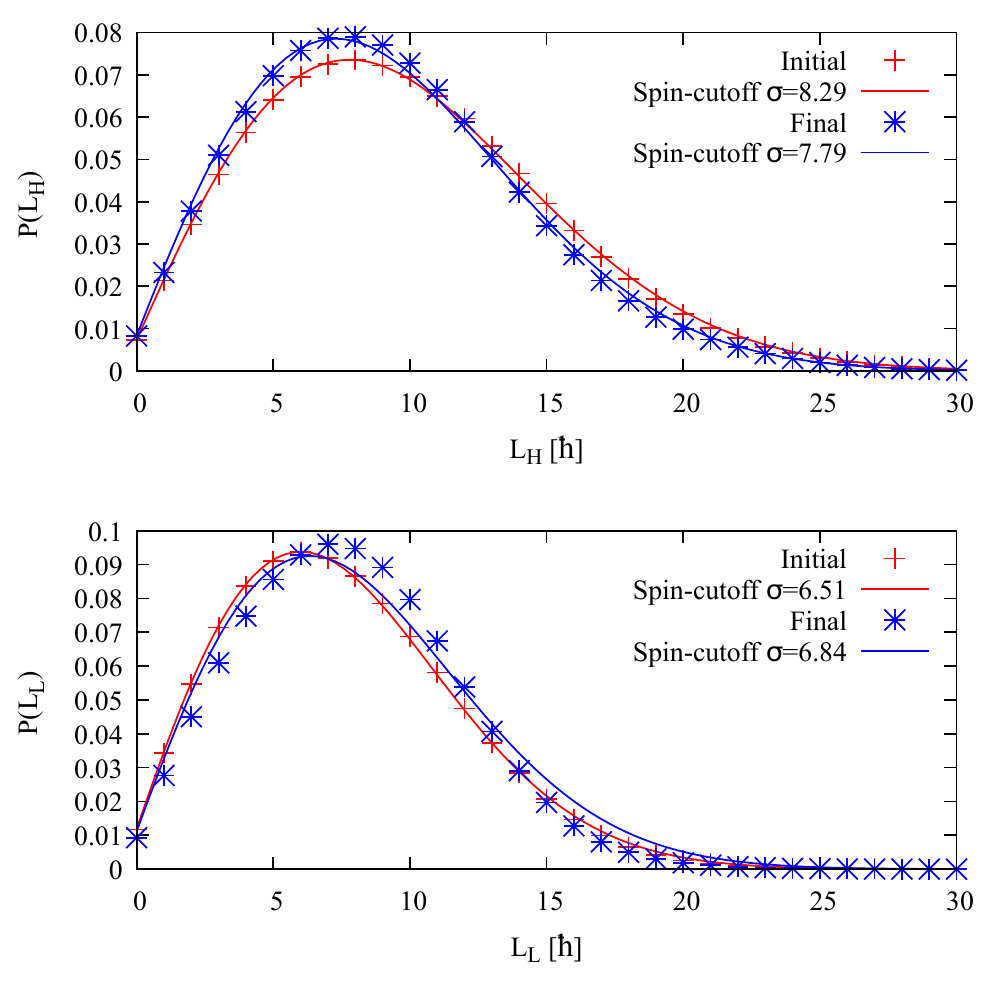}
\caption{ (Color online) Initial and final spin distributions of 
the heavy (top) and light (bottom) fragments in the fission
reaction $^{240}$Pu $\rightarrow$ $^{144}$Ba+$^{96}$Sr. Solid
lines show the comparison to Eq. \eqref{eq:Beth}.
 } 
\label{fig:fit_tfinal_l1_l2_1D}
\end{figure}

The case where both fragments are deformed presents a more  rich dynamics with the possibility of correlation between the two angular momenta. But first, we determine the angular momentum distribution of each fragment.
  
 The angular momenta of the fragments are shown in Fig.  \ref{fig:fit_tfinal_l1_l2_1D} at the initial distance and at the final one. The angular momentum distribution seen in the light fragment is found to be smaller than in the heaviest one. It will be shown in Sec. \ref{sec:oct} that it is due to the octupole deformation in the heavy fragment.

To see if the  distribution obtained by the collective Hamiltonian model deviates from the Bethe formulae,
the figure shows a comparison between the spin-cutoff formulae, 
 \begin{align}
 P(L) \propto (2L+1) e^\frac{-L(L+1)}{(2 \sigma^2)}, \label{eq:Beth}
 \end{align}
 and the initial and final spin distribution. 
 A good agreement is found with the formulae for the initial distribution, but dynamics change the shape of the distribution. Especially for the light fragment which is more affected by the Coulomb torque.

\section{Angular momentum Correlations}
\label{sec:correlation}

\subsection{Correlation between the magnitudes of the angular momenta}
\label{sec:corr_mag}
\begin{figure}[!h]
\centering
\includegraphics[width=.99\linewidth, keepaspectratio]{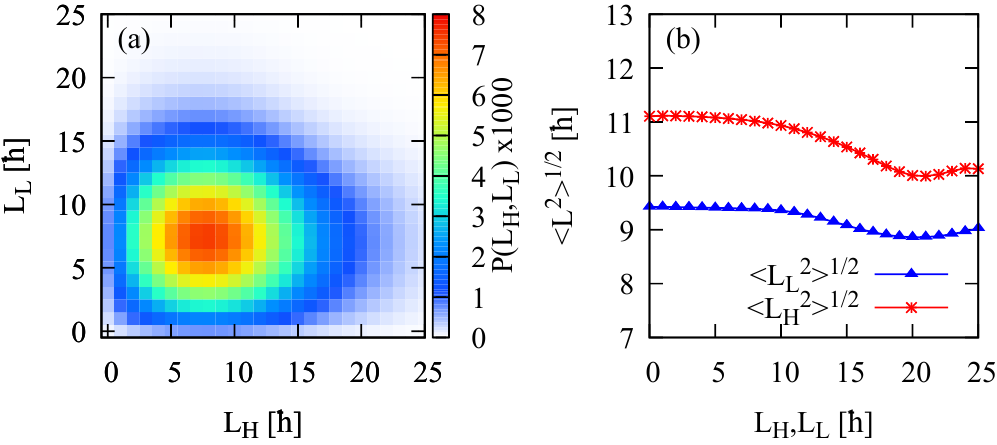}
\caption{ (Color online) (a) Distribution of the angular momentum of the heavy and light fragment. (b) Average angular momentum of one of the fragments as a function of the angular momentum of the other fragment.
 } 
\label{fig:fig_corr_amplit}
\end{figure}

The  reaction $^{240}$Pu $\rightarrow$ $^{144}$Ba+$^{96}$Sr
is interesting for examining  the correlations between the fragment's
angular momenta since both are deformed.  
The two-dimensional probability distribution of angular momentum
magnitudes $P(L_H,L_L)=\sum_m |\Psi(L_H,L_L,m)|^2$
is shown in panel (a) of Fig. \ref{fig:fig_corr_amplit}.  
There is hardly any correlation
between the two magnitudes.  For a quantitative measure, 
panel (b) shows  the average angular momentum of one of the
fragments as a function of the other,
\begin{equation}
\begin{aligned}
<\hat L_L^2>_{L_H}&=\frac{\sum_{L_L} L_L(L_L+1) P(L_H,L_L) }{ \sum_{L_L} P(L_H,L_L) }, \\
<\hat L_H^2>_{L_L}&=\frac{\sum_{L_H} L_H(L_H+1) P(L_H,L_L) }{ \sum_{L_H} P(L_H,L_L)}.
\end{aligned}
\end{equation}
One sees  only a small
correlation  between the magnitude of the angular momentum of the fragments.  The angular momentum of the heavy
fragment varies by about one unit depending on the light
fragment.  This result is compatible with the experimental data
of ref.~\cite{Wil21}.

\subsection{Correlation between the orientations of the angular momenta}

\begin{figure}[t]
\centering
\includegraphics[width=.79\linewidth, keepaspectratio]{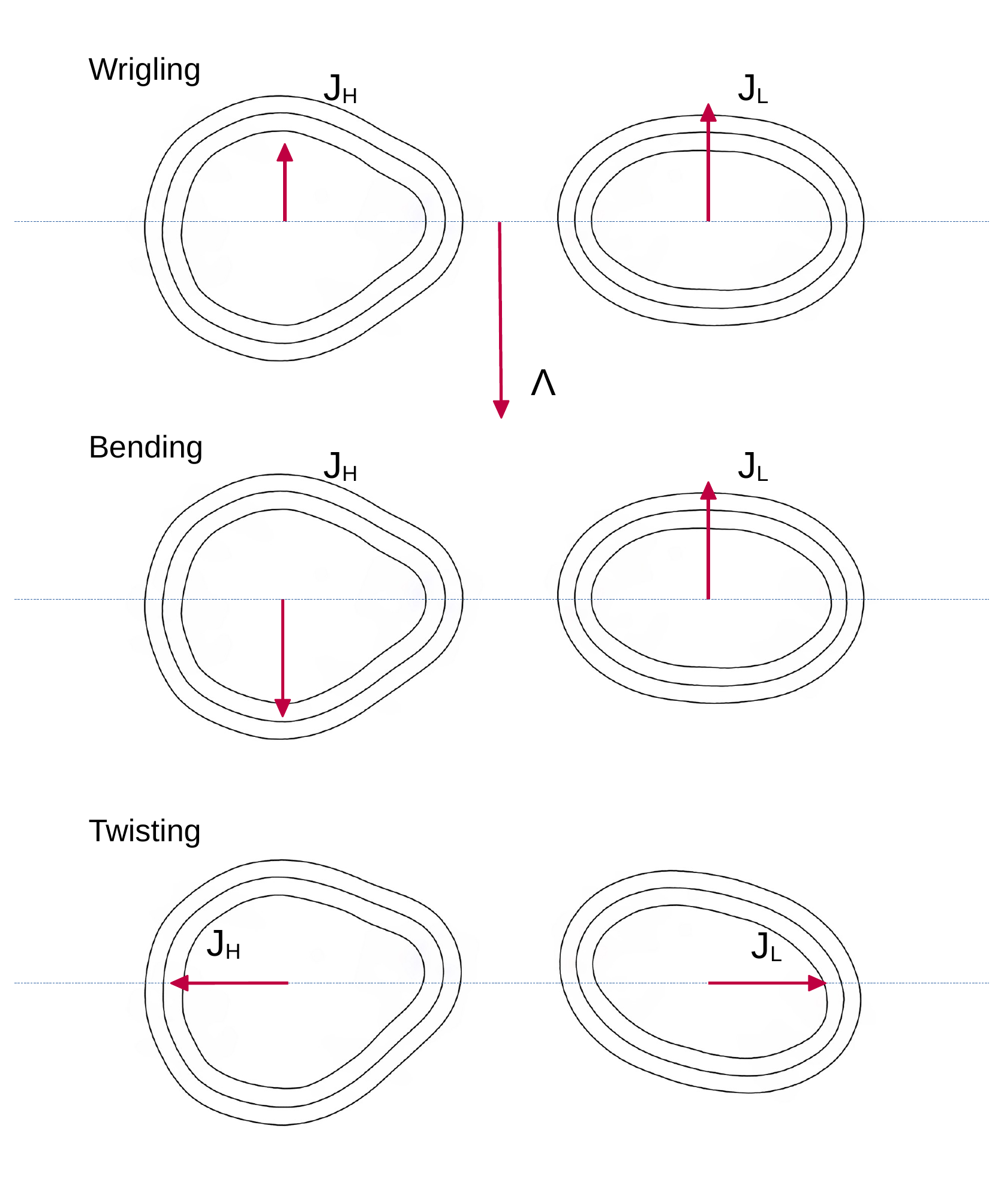}
\caption{ (Color online)  Schematic description of the type of correlation between the two fission fragments' angular momenta. The red arrow represents the angular momentum of the fragments.
 } 
\label{fig:explain_modes}
\end{figure}

We describe here the different types of correlations, based on the nomenclature of the modes. These modes have been introduced in models that assume the fragments rotate in a strongly correlated way as in a rigid classical model. However, in the present model, we describe the system with a collective wave packet that describes soft correlations between the orientation of the two fragments' angular momenta. The three types of correlations described in the following are shown in Fig. \ref{fig:explain_modes}. The correlations of the orientation of the angular momenta are described as bending when the spins are in the same direction and in the plane perpendicular to fission, wriggling when they are in opposite directions, and twisting when the angular momenta are aligned in the fission axis.   Tilting is ignored in this figure since it is a forbidden mode in our model
due to the assumption that the total angular momentum is zero. In practice, our model mixes the three kinds of correlations with a wave packet describing the directions of the angular momenta.

\begin{figure}[h]
\centering
\includegraphics[width=.99\linewidth, keepaspectratio]{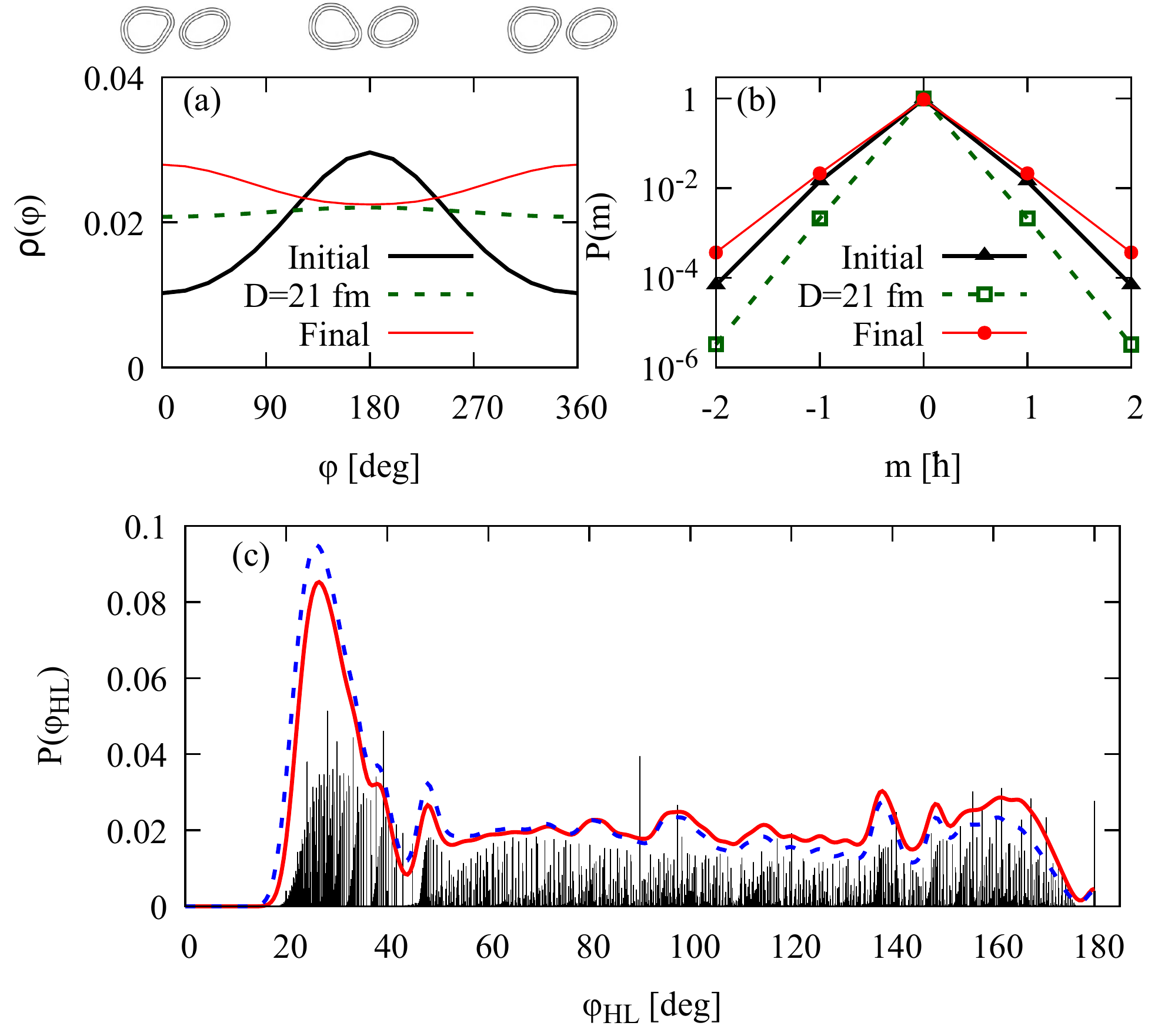}
\caption{ (Color online) Angular momentum correlations in the fission reaction (a) Distribution of the relative azimuthal angle $\varphi$ (describing the spatial orientation of the fragments), at the initial time, scission, and large distance. The corresponding configuration is shown above the figure. (b) Distribution of the projection of the angular momentum on the fission axis. (c) Final distribution (unnormalized) in an arbitrary unit of the opening angle between angular momenta computed as Eq. \eqref{eq:openang} (black bar). The distribution averaged with a 2$^{\circ}$ wide Gaussian function is shown with a red line. The dashed blue line shows the initial distribution.
 } 
\label{fig:Angul_correl}
\end{figure}

Although the magnitudes of the angular momenta are uncorrelated (see Sec. \ref{sec:corr_mag}),
there is a substantial correlation between their directions.  As 
early studies have shown~\cite{Wil72}, both
angular momenta are largely perpendicular to the fission axis.
This implies that $m=0$ is favored in the angular momentum representation;
 in the present model the probability of $m \neq 0$ is of the
order of 1\% (see Fig. \ref{fig:Angul_correl}(b)).  
This affects the
angular correlation with respect to
the relative azimuthal angle
between the two fragments $\varphi=\varphi_H-\varphi_L$.  

Although the non-zero m components are small, there are visible correlations
 in the black curve in Fig. 
\ref{fig:Angul_correl}(a). The configuration with relative azimuthal angle $\varphi=180$$^{\circ}$ is two times more probable than the one with $\varphi=0$$^{\circ}$.  

As can be seen in Fig.  \ref{fig:pot_fct_theta_Ba_Sr} in Appendix \ref{sec:pot3D}, the lowest energy
configuration is obtained for $\varphi=180$$^{\circ}$ which corresponds to a
shape in V.    
This correlation arises
from the nuclear interaction between the two fragments.  At the
scission point, the correlation is suppressed, resulting in the
disappearance of non-zero $m$ components.  However, during the
separation phase, the magnitudes of the non-zero $m$ components
increase (See Fig.  \ref{fig:Angul_correl}(b)) due to the
coupling induced by the $\Lambda^2$ term in the Hamiltonian (Eq. 
\eqref{eq:Hamilt}).  Despite this, the $m=0$ states continue to
largely dominate, which is consistent with the experimental data
from Ref.~\cite{Wil72}.

\subsection{Opening angle}

Recently, the distribution of opening angle $\varphi_{HL}$
between the fragments' angular momentum vectors has been much
discussed.  This distribution differs in the various models
\cite{Ran21,Ran22,Bul22,Ran22b,Bul22b}.  We define the
distribution as the operator with diagonal elements in the
$| L_H, L_L, \Lambda , M \rangle$ basis
\begin{equation}
\begin{aligned}
 \varphi_{HL} = \arccos\left( \frac{ \Lambda (\Lambda +1) - L_H (L_H +1) - L_L(L_L+1) }{2 \sqrt{L_H (L_H +1) L_L (L_L +1) } } \right) \label{eq:openang}
\end{aligned}
\end{equation}
and zero off-diagonal elements.  Its probability distribution
is  obtained from the distribution of states $ |L_H, L_L, \Lambda \rangle $. 
\begin{equation}
\begin{aligned}
P(\varphi_{HL}) &= \sum_{ \substack{\Lambda,L_H>0, \\L_L>0 }} \delta(\varphi_{HL}-\varphi_{HL}(\Lambda,L_H,L_L)) \nonumber \\ &\left| \sum_m (L_H,m,L_L,-m | \Lambda, 0) \Psi(L_H,m,L_L,-m) \right|^2. \label{eq:distr_open}
\end{aligned}
\end{equation}

This formula is similar to the semi-classical formula  in
Ref.~\cite[Eq. (4)]{Bul22}.  To replace $\sqrt{L_{H,L} (L_{H,L} +1)}$ by $L_{H,L} +1/2$ leads to a very similar
distribution.  Note that the value of $\varphi_{HL}$ is fixed
by the quantum numbers of component  $| L_H, L_L, \Lambda
\rangle$, so the probability distribution is a set of discrete
spikes.
Fig. 
\ref{fig:Angul_correl} shows that distribution.  It has a large
concentration of strength near  25$^{\circ}$ associated with the states in which
$\Lambda=L_H+L_L$.  The average angle is found to be
86.7$^{\circ}$ (78.4$^{\circ}$ for the initial distribution). 
None of the models described in ref. 
\cite{Ran21,Ran22,Bul22,Ran22b} has this feature.  In the
language of the collective vibrational models, these values
would correspond to the presence of a wriggling mode.
This distribution is close to the one found in time-dependent density functional theory \cite{Sca23}, with the difference that the distribution of K in the microscopic approach diminishes the peak at small opening angles.
 
 \subsection{Correlation in a direction perpendicular to the fission axis}
 
\begin{figure}[!h]
\centering
\includegraphics[width=.99\linewidth, keepaspectratio]{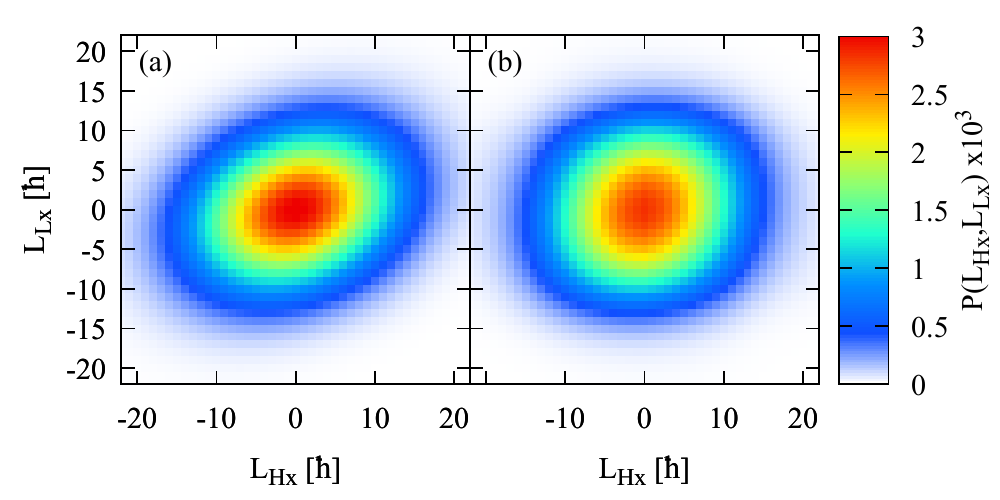}
\caption{ (Color online) 
Correlations between the projections of the angular momentum on the x-axis at initial (a) and final (b) times.
 } 
\label{fig:cor_dirx}
\end{figure}

To determine the degree of alignment of the angular momenta,
the correlated wave function is projected in the basis $| L_H,
L_{Hx}, L_L, L_{Lx} \rangle$.  This enables the calculation of
the correlation between the angular momentum projected on a
transverse axis (called "x"):
\begin{equation}
\begin{aligned}
 P(L_{Hx},L_{Lx}) =\sum_{L_H,L_L} |\Psi(L_H, L_{Hx}, L_L, L_{Lx})|^2.
\end{aligned}
\end{equation}
To compute the correlation between the spin in the x direction, the angular momentum basis $L,m_z$ is turned by 90 degrees in the y direction.
 The coefficients in the turned basis are, 
 \begin{widetext}
 \begin{align}
\Psi_x (L_H, L_{Hx}, L_L, L_{Lx}  ) = \sum_{ L_{Lz}, L_{Hz}}  {\cal D}^{J}_{L_{Hx},L_{Hz}} (0,\pi/2,0)  {\cal D}^{J}_{L_{Lx},L_{Lz}}(0,\pi/2,0)  \Psi (L_H, L_{Hz}, L_L, L_{Lz}  )
 \end{align}
  \end{widetext}
with Wigner D-function ${\cal D}^{J}_{M,M'}(\alpha,\beta,\gamma)$.
The two-dimensional probability plot is shown in Fig.  \ref{fig:cor_dirx}.  
The angular momenta are  correlated initially, but the 
correlation is much reduced in the final state.
The correlation coefficient
$r_{L_{Hx},L_{Lx}}=\frac{{\rm
Cov}(L_{Hx},L_{Lx})}{\sigma_{L_{Hx}} \sigma_{L_{Lx}}}$  
is 0.22 initially and 0.06 at the final time.  The positive sign
of the coefficient indicates the
presence of a wriggling correlation type as opposed to a bending type in the vibrational model.
 The probability that both projections on the x-axis have the same sign gives the population of the wriggling mode and is found to
be 0.58 initially and 0.53 at the end of the evolution.

This may seem to  contradict the $\varphi$ correlation of the
rotors' axes that peak at zero as in a bending mode.  
The reason for this apparent contradiction is as
follows: while the potential may favor a V-shaped configuration,
the angular momenta of the fragments actually originate from
their zero-point motion.  Since the confining potential is
stiffer in the direction of $\theta_H-\theta_L$ than in the
direction of $\theta_H+\theta_L$, the wriggling mode ends up
dominating.  This result highlights the importance of
considering the collective Hamiltonian as a function of all the
orientation angles in understanding the angular momentum modes
of fission fragments.

\begin{table}[h]
\caption{ Average spin$\langle L^2 \rangle^{\frac12}$ in unit of $\hbar$ for the 3 fission fragments at scission ($D=21$ fm) and at large distances. The last two columns show the same quantity with an MOI divided by 2. }
\begin{tabular}{|c|c|c|c|c|}
\hline
Nucleus & Scission & Final & Scission ($I_{\frac{1}{2}}$) & Final ($I_{\frac{1}{2}}$) \\
\hline
\hline
 $^{108}$Ru & 9.28 & 12.31 & 7.24 & 10.38 \\
$^{144}$Ba & 10.04 & 10.95 & 7.70 & 8.66 \\
$^{96}$Sr & 7.74 & 9.30 & 6.03 & 7.62 \\
\hline
\end{tabular}
\label{tab:res}
\end{table}

\section{Parameter sensitivities}
\subsection{Moment of inertia}
\label{sec:MOI}
\begin{figure}[!h]
\centering
\includegraphics[width=.99\linewidth, keepaspectratio]{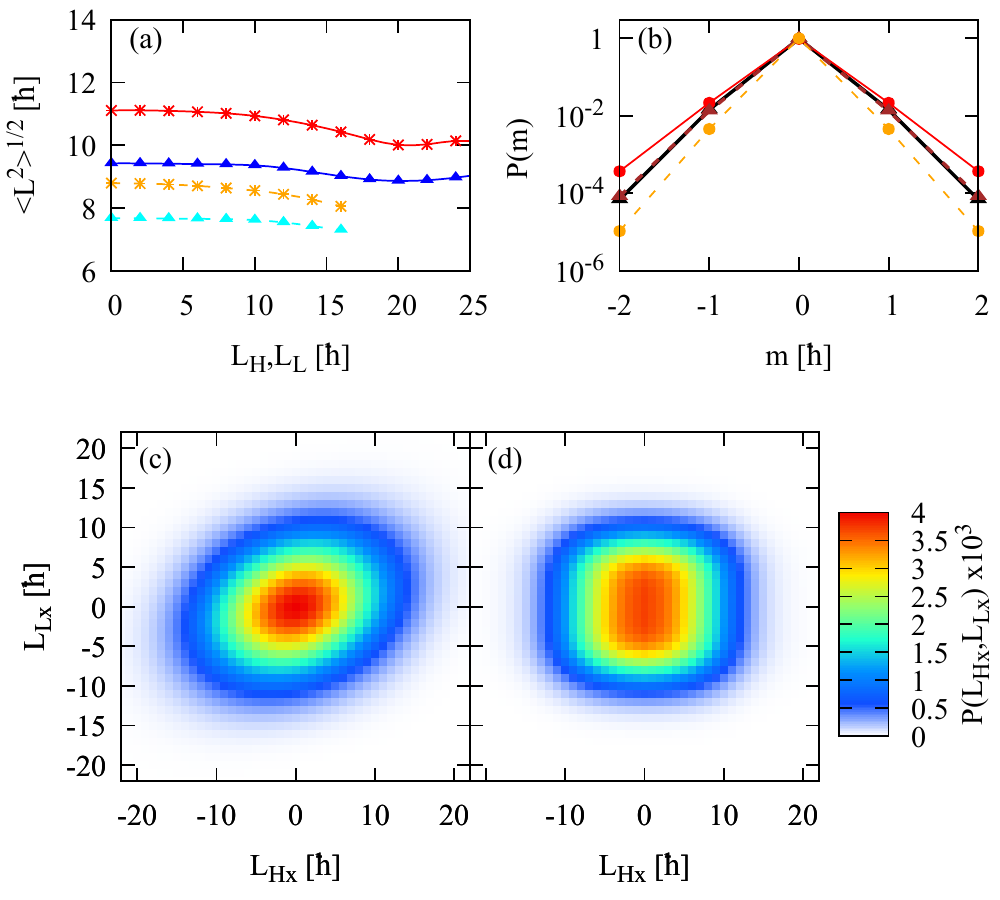}
\caption{ (Color online) Effect of the moment of inertia on the angular
momentum correlation in the fission reaction $^{240}$Pu $\rightarrow$ $^{144}$Ba+$^{96}$Sr.  (a) Average angular momentum of one (heavy fragment is shown with crosses and light fragment with triangles) of the fragments as a function of the angular momentum of the other fragment for the rigid moment of inertia (solid line) and half the moment of inertia (dashed line). (b) Distribution of $m$ at the initial (triangle) and final time (dots) for the rigid moment of inertia (solid line) and half the moment of inertia (dashed line). Panels (c) and (d) show the correlations between the angular momentum projected on the x-axis respectively at the initial and final time in the case of the half rigid moment of inertia.
 }
\label{fig:effect_MOI_4D}
\end{figure}

So far, we have assumed a rigid-body moment of inertia (MOI) in Eq. \eqref{eq:Hamilt}.
The empirical moments are smaller by as much as a factor of two
due to pairing effects~\cite{Boh75}.  
To estimate the impact of a
different MOI, calculations have been performed
with the MOI divided by two.  Results are shown in Fig. \ref{fig:effect_MOI_4D} and
Table  \ref{tab:res}.  We see that a smaller MOI
reduces the average spin in both the initial and final
distribution by about $2\hbar$.  

 \begin{figure}[!h]
\centering
\includegraphics[width=.99\linewidth, keepaspectratio]{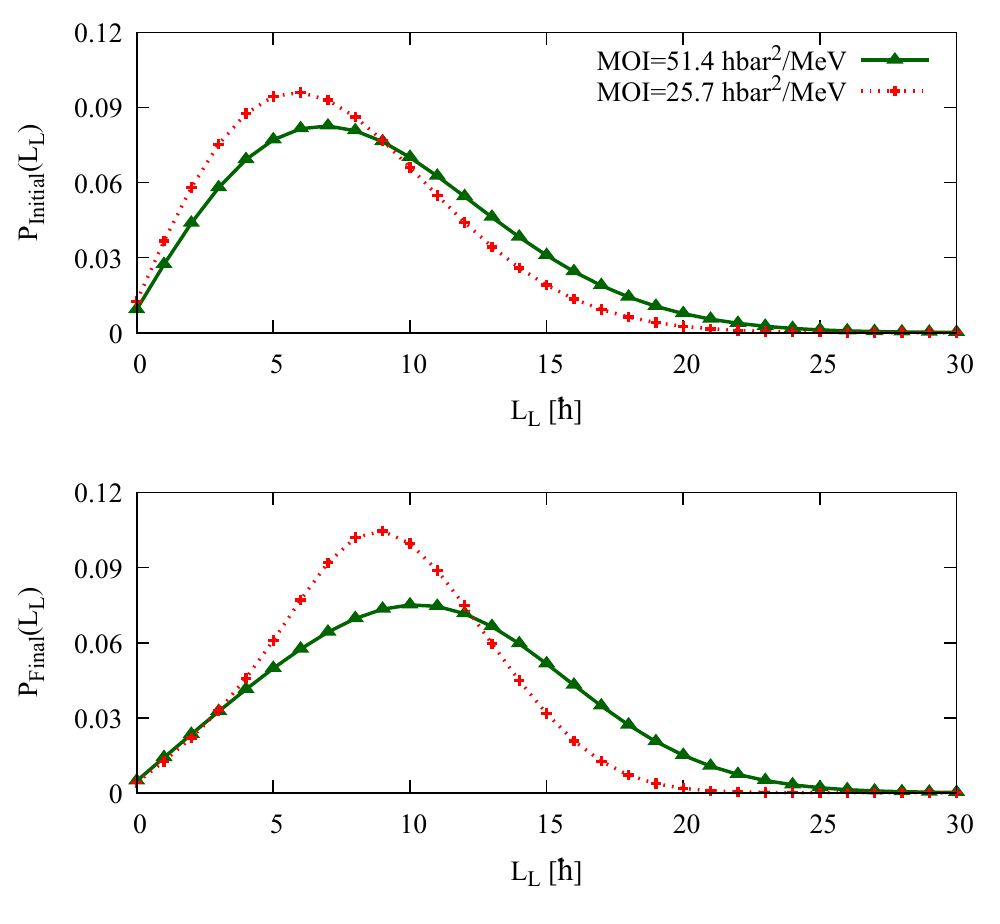}
\caption{ (Color online) Effect of the moment of inertia in the initial 
(top) and final (bottom) angular momentum distribution in the $^{108}$Ru
fission fragment.}
 
\label{fig:effect_MOI_1D}
\end{figure}

The reduction occurs already in the initial angular momentum associated
with the starting wave function. This is in accord with the qualitative argument in Ref. \cite{Bon07} which  discusses the angular localization
produced by an attractive pocket in the nuclear potential between the
post-scission fragments.  The average angular
momentum was predicted to scale as $I^{1/4}$;  the numerical results in
Table \ref{tab:res} are close to that with a slightly stronger dependence.

The angular momentum generated in the post-scission evolution ranges from 1
to 3 units, and is quite independent of the MOI.  An analysis of the quantum
mechanics of that process is given in Appendix \ref{sec:torque}, showing that it
depends strongly on the dephasing of the components of the wave function in
the angular-momentum representation.  While the inertial dynamics can
contribute to the dephasing, other dephasing mechanisms apparently
dominate.

The average final angular momentum for the Hamiltonian with 
a reduced MOI is in the range 7.5-10.5$\hbar$,
significantly higher than the experimental values for 
comparable nuclei~\cite{Wil21}. This could be due to model limitations or angular momentum carried by emitted neutrons~\cite{Ste21,Mar21}.
 
\subsection{Choice of $D_0$}

The choice of the point to start the Schr\"odinger evolution is not well
determined, since the dynamics of the neck elongation and thinning are beyond
the scope of the model. We chose $D_0 = 14$ fm as a point where the two
proto-fragments overlap but the scission has proceeded enough to where the
mass splitting between them is clear.  To get an idea of the model
uncertainty of this parameter we have also analyzed cases with larger values
of $D_0$.  
\begin{figure}[!h]
\centering
\includegraphics[width=.99\linewidth, keepaspectratio]{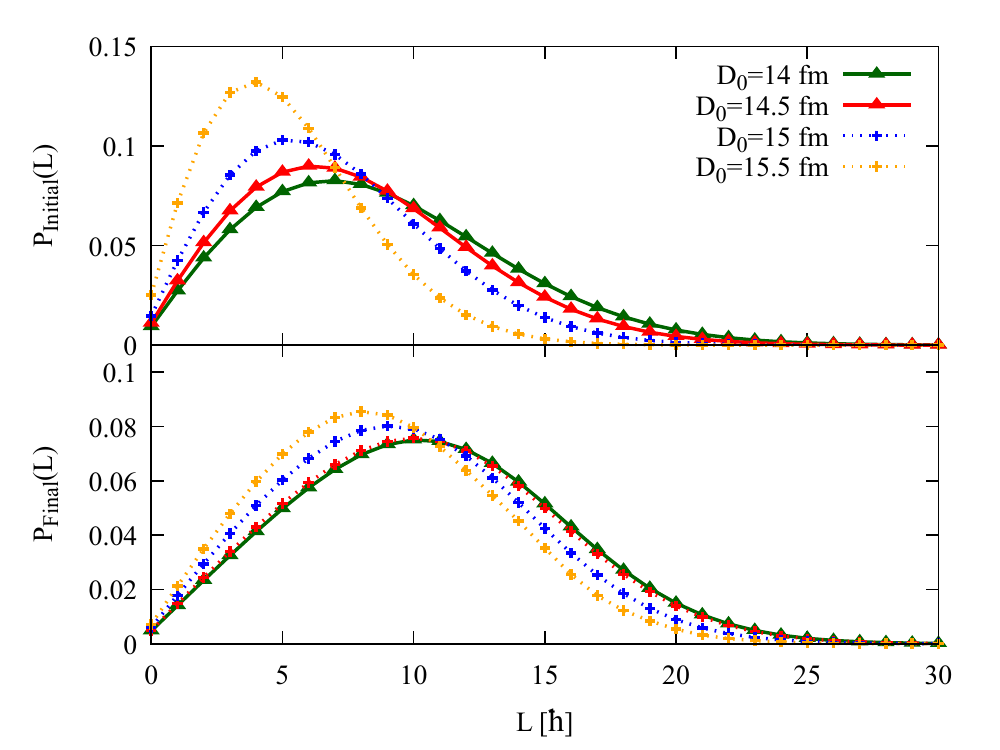}
\caption{ (Color online) Distribution of $L_L$ at t=0 (top) and final time (2nd panel) for different values of $D_0$ in the $^{132}$Sn+$^{108}$Ru case. } 
\label{fig:effect_Dini_1D}
\end{figure} 
Initial and final distributions for the range $D_0 = 14-15.5$ are shown 
in Fig. \ref{fig:effect_Dini_1D}).  There is quite a strong dependence in 
the initial distributions, but the differences are attenuated in the final
distributions.   The main reason is the slow
overdamped evolution before scission which makes the evolution
quasi-adiabatic and so less dependent on the initial conditions. 
Nevertheless, the difference in the final angular momenta for 
$D_0=14$ and 15.5 fm are 20\%.  We may be taken as the model uncertainty
at a quantitative level.

\section{Impact of octupole deformation}
\label{sec:octupole}

\label{sec:oct}

\begin{figure}[!h]
\centering
\includegraphics[width=.99\linewidth, keepaspectratio]{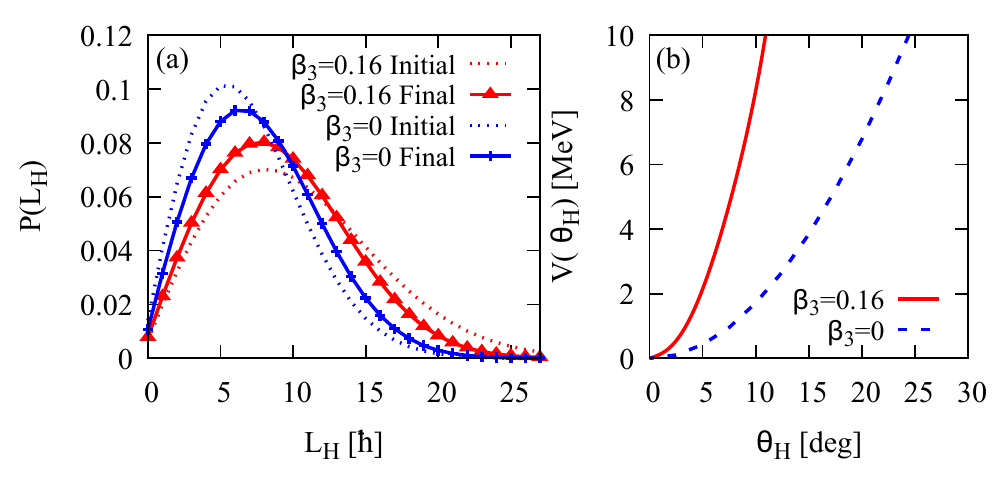}
\caption{ (Color online) Comparison with and without taking into account the octupole deformation. (a) Distribution of angular momentum at the initial and final time (b) Angular potential at a distance $D$=14 fm. This calculation is done in one dimension, assuming a fixed light fragment with $\theta_L=0$.
 } 
\label{fig:effect_oct}
\end{figure}

There is one 
unexpected qualitative observation that can be explained
within the FHF approach. Namely, the experimental results
generally favor higher angular momentum in a more strongly
deformed fragment, but the data in Ref.~\cite{Wil21} for
$A_H\simeq144$ and $A_L\simeq96$ have the opposite behavior.
Although the quadrupole deformation of $^{144}$Ba is small,
that nucleus has a large pear-shaped deformation (octupole)
at the scission point which affects its angular momentum
content.  This may be seen in Fig.  \ref{fig:effect_oct} 
which shows the angular momentum distribution with and without taking into account the
pear-shaped deformation. Both initial and final angular momenta are increased when octupole deformation is present. More precisely, the final average angular momentum increases from $\langle L_H^2
\rangle^{1/2}$ = 9.3 to 10.9 $\hbar$ when the calculation includes octupole deformation.  

The effect is explained by the difference in stiffness of the angular
potential as seen on panel (b).  As shown in Ref.~\cite{Sca18},
the octupole deformation in the fission fragment minimizes the
energy at the scission which promotes the production of fission
fragments with an octupole deformation in their ground
state.  Here, the pear shape maximizes the nuclear interaction
between the fragments at $\theta_H$=0 and creates a stiffer
potential at small angles.  The stiffer potential induces a
larger initial angular momentum.

\section{Conclusion}

The model presented here addresses the following questions regarding
the collective contribution to the angular
momentum generated in fission:\\
--What is the angular momentum distribution in the fission fragments?\\
--What correlations exist between the angular momenta of
the two fragments?\\
--How does it depend on the Hamiltonian dynamics in the various states of
the scission process?\\
In presenting the model conclusions it should be remembered that the model
ignores the neck dynamics and cannot provide complete answers.  
First of all, the nuclei are considered
to be cold in the sense that quasi-particle excitations are not included.
These excitations also carry angular momentum and will augment it.  Also,
those excitations may couple to the collective component, allowing the
Coulomb moments to damp out with time.  Obviously, this mechanism
would decrease the post-scission contribution.

The most easily grasped observable in this study is the total angular
momentum generated in each fission fragment.  We see from the examples
that it depends strongly on the fragment deformation and most of the
angular momentum is present at an early stage, arising from the quantum
uncertainty between angular momentum and orientation angle.  
While normally a large deformation would permit
larger angular momentum, a surprising finding is that the
lightly deformed $^{144}$Ba emerges with a higher angular momentum  
than its partner fragment due to an octupole
deformation.  A second finding is that the angular momentum gained 
in the post-scission evolution is of the order of 1-3~$\hbar$.  The
gain  is
insensitive to the MOI of the fragments, unlike the initial angular
momentum which decreases with the MOI.

Our analysis of the correlations between
the angular momenta of the two fragments
should provide a deeper insight into the effect of the potential at
the scission point.  There is hardly any correlation between the
magnitudes of the angular momenta in the two fragments, in agreement
with the experimental data of Ref. [3]. For the angular correlation
between them, the potential favors  V-shaped
configurations that would correspond to bending mode vibrations and produce
a negative angular correlation.
We find that the correlations are small but positive, corresponding to
a wriggling mode of vibration.


There are several ways the treatment of the neck-breaking could be improved. The Density-Constrained Hartree-Fock (Bogoliubov)~\cite{Cus85,Uma85,Sca19,God22} that contrarily to the FHF approximation does not violate the Pauli principle.
 Scission configurations could also be obtained from constrained calculations~\cite{Ber90,War02,Han21} or deduced from TDHF evolution~\cite{Was09}. 
 The MOI of fragments could be obtained from a microscopic Hamiltonian~\cite{Was21}.
Finally, the rotation of the fragments is not completely
collective~\cite{Sca22}, which raises the question of how to
simultaneously treat the collective and single-particle degrees
of freedom~\cite{Arv81,Neg82,Reg19,Ver20,Bul19,Li23,Mav23}.

\medskip

\begin{acknowledgments}
We thank the organizers A.  Bulgac, J.  Randrup,
I.  Stetcu, and J.N.  Wilson as well as all the participants of the workshop on fission fragment
angular momenta which led to interesting and inspiring
discussions.
 And a particular thanks to Lee Sobotka for discussions at an early stage of the project.
The funding from the US DOE, Office of Science,
Grant No.  DE-FG02-97ER41014 is greatly appreciated.  This
research used resources of the Oak Ridge Leadership Computing
Facility, which is a U.S.  DOE Office of Science User Facility
supported under Contract No.  DE-AC05-00OR22725.
\end{acknowledgments}

\appendix

\section{Frozen Hartree-Fock potential, 1D case}
\label{sec:pot1D}

The static potential has been obtained using the Frozen Hartree-Fock method on a grid of values of $D$ and $\theta$.
Solving the collective Hamiltonian model requires a finer grid. Then, a fit has been done of the FHF potential with the following form, with a separation of Nucleus-Nucleus interaction $V_{NN}( \theta, D) $ and the Coulomb interaction $V_{C}( \theta, D)$,
  \begin{align}
V_{NN}( \theta, D) &= \frac{ V_0(D) }{ 1 + \exp^{ \frac{-( | \sin(\theta)|^{0.5} - t_0(D))}{ a} } }, \label{1D_NN_pot} \\
V_{C}( \theta, D) &= \frac{e^2 Z_H}{2D^3} P_2(\cos(\theta))  Q_2^C  \nonumber \\ &+  \frac{e^2 Z_H}{2D^5} P_4(\cos(\theta))  Q_4^C,  \label{eq:pot_coul}
\end{align}
with,
\begin{align}
V_0(D) &= \exp(V_{0a} D^2 + V_{0b} D + V_{0c} ), \\
t_0(D) &= t_{0a} D^2 + t_{0b} D + t_{0c}. 
\end{align}

 \begin{figure}[!h]
\centering
\includegraphics[width=.99\linewidth, keepaspectratio]{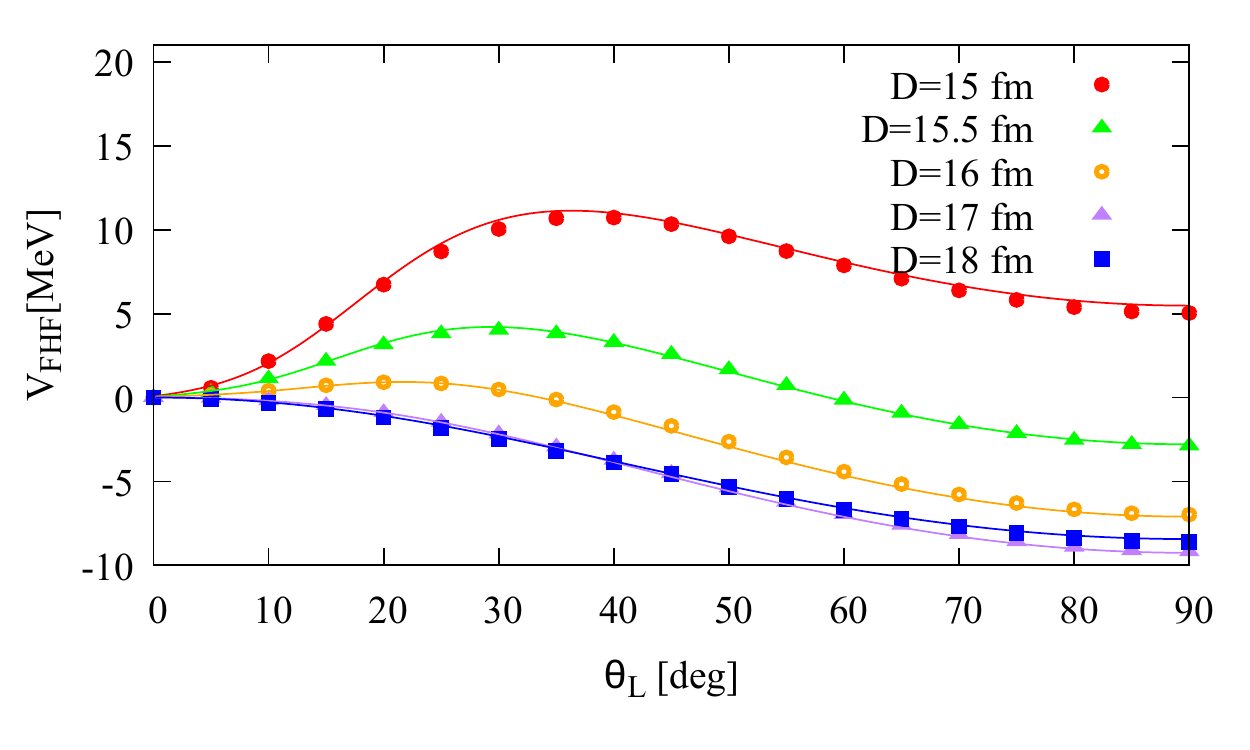}
\caption{ (Color online) Comparison between the Frozen Hartree-Fock potential (symbol) and the fitted potential eq. \eqref{1D_NN_pot} (lines) in the reaction $^{240}$Pu $\rightarrow$ $^{132}$Sn+$^{108}$Ru.
 } 
\label{fig:pot_fct_theta_Sn_Ru}
\end{figure} 

\begin{table}[!h]
\caption{ }
\begin{tabular}{c|c|c|c|}
\hline
\hline
 Variable & Sn+Ru & Ba(no oct.)+Sr &Ba(oct.)+Sr \\ %
\hline
$a$ & 0.1 & 0.13 & 0.106 \\
$V_{0a}$ [MeV] & -0.1776 & -0.1604 & -0.1548 \\
$V_{0b}$ [MeV]  & 4.117 & 3.660 & 3.673\\
$V_{0c}$ & -18.76 &-15.87 & -16.56\\
$t_{0a}$ & 0.00609 & 0. & 0.00393\\
$t_{0b}$ & -0.2174 & 0. & -0.143 \\
$t_{0c}$ & 2.501 & 0.82 & 1.903 \\
$Q_2^{C}$ [fm$^3$] & 614.5 & 276.14 & 299.88 \\
$Q_4^{C}$ [fm$^5$] & 12483.4 & 3907.63 & 23007.7 \\
\hline
\end{tabular}
\label{tab:val1Dpot}
\end{table}

The values in the case of $^{132}$Sn+$^{108}$Ru and simplified
$^{144}$Ba+$^{96}$Sr with fixed $^{96}$Sr with and without octupole
deformation are shown on Table \ref{tab:val1Dpot}.  To test the validity of
that parametrization a comparison between the FHF and the fitted potential
is shown in Fig.  \ref{fig:pot_fct_theta_Sn_Ru}.

\section{Frozen Hartree-Fock potential, 3D case}

\label{sec:pot3D}

\begin{figure}[!h]
\centering
\includegraphics[width=.99\linewidth, keepaspectratio]{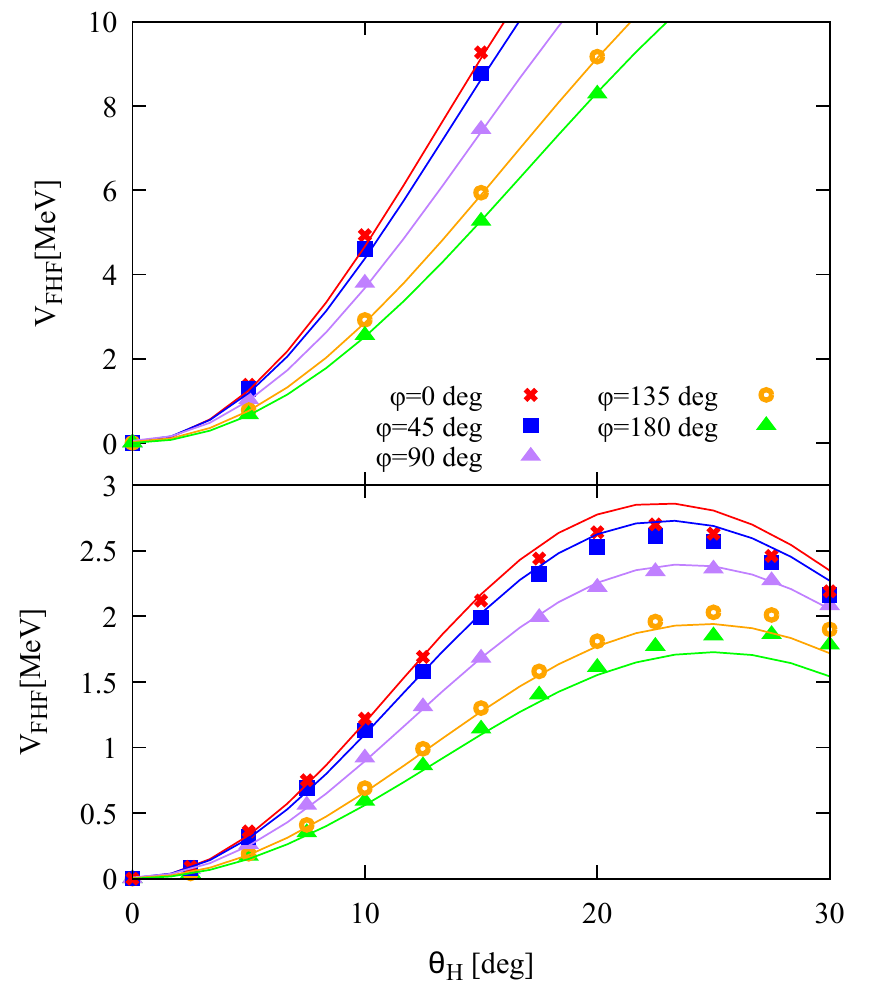}
\caption{ (Color online) Comparison between the Frozen Hartree-Fock potential (symbol) and the fitted potential eq. \eqref{1D_NN_pot} (lines) for $D$=15.5 fm and $D$=16.5 fm. $\theta_L$=$\theta_H$=25 degrees.
 } 
\label{fig:pot_fct_theta_Ba_Sr}
\end{figure}

In the case of the $^{144}$Ba+$^{96}$Sr the same angular grid has been
used for the FHF calculation and in the collective Hamiltonian model. 
Rather than publishing the very large table of entries for $\hat V$
 we provide in
the Supplementary Material~\cite{suppl} a Fortran program to evaluate the
four-dimensional internuclear potential. It has been checked that this
potential leads to the same results as the original FHF one (See Fig. 
\ref{fig:pot_fct_theta_Ba_Sr}).  In the Schr\"odinger code to evolve the wave
function, nuclear matrix elements are interpolated as a function of $D$ from
computed values on a grid of spacing of 0.5 fm 
between $D=14$ and $D$=17 fm and
1 fm between $D=$17 and $D=20$ fm. For the other values of $D$
the matrix elements are interpolated assuming an exponential evolution from
the two closest existing matrix elements.

The Coulomb potential is parametrized as  
\begin{align}
 &V_{C}( \theta_H,\theta_L, D) \nonumber \\&= \sum_{l=2}^{4} \frac{e^2}{2} \left( \frac{Z_L}{D^{l+1}} P_l(\theta_H) Q_{lH}^C + \frac{Z_H}{D^{l+1}} P_l(\theta_L) Q_{lL}^C  \right).
\end{align}

The Coulomb moments are obtained by fitting the FHF Coulomb potential
and are given in Tab.  \ref{tab:Coul}.  These are close to the integrated
values of Eq.  \eqref{eq:mom}.

\begin{table}[h]
\caption{ Parameters of the Coulomb moments}
\begin{tabular}{|c|c|}
\hline
$Q_{2H}^C$ & 299.88  fm$^2$ \\
$Q_{2L}^C$ &  352.96  fm$^2$ \\
$Q_{3H}^C$ & 1453.96  fm$^3$ \\
$Q_{3L}^C$ & 0  fm$^3$ \\
$Q_{4H}^C$ & 9153.18  fm$^4$ \\
$Q_{4L}^C$ &4849.29  fm$^4$ \\
\hline
\end{tabular}
\label{tab:Coul}
\end{table}

\section{Angular momentum and torque}
\label{sec:torque}
The angular momentum $L$ in a system is changed by a torque 
in both classical and quantum mechanics, but the torque induced by
the post-scission Coulomb field on the fragment's angular momentum 
is somewhat subtle.  In this section, we derive analytic formulas for
the time dependence of $L$ to see why the net post-scission increase is quite small and dependent on the moment of inertia.
The analysis proceeds with a generalization of Ehrenfest's  theorem to 
relate the expectation value of the torque to the rate of change
of angular momentum,
\begin{align}
\frac{d}{dt} \langle \hat L \rangle = \frac{1}{i \hbar} \langle [\hat L, \hat H] \rangle.
\end{align}
Here $\hat L$ and $\hat H$ are operator in the space of orientation angles
$\theta$ and $\varphi$, and the brackets denote expectation values with respect to the 
instantaneous wave function. The commutator on the right  is the
quantum mechanical torque.  The Hamiltonian $\hat H$ is given by
\begin{align}
\hat H = \frac{1}{ 2 I} \hat L^2 + a P_2 (\cos
(\theta))
\end{align}
where $ a =  Z Q_f/(D(t))^3$ in Eq. (\ref{eq:pot_coul}).  
The Hamiltonian  commutator with $\hat L_x$ evaluates to
\begin{align}
[\hat L_x, \hat H]  = -3i\, \cos(\theta) \sin(\theta) \sin(\varphi). 
\end{align}
Notice that the torque is negative in the range $0 < \theta < \pi/2$,
 positive in the range $\pi/2 < \theta < \pi$, and zero at the endpoints
of these intervals.  This is just what one expects for a classical torque
about the $x$-axis from a force derived from a potential that is symmetric
under reflection in the $xy$ plane.  For example, the  wave function
$\Psi(\theta,\varphi) \sim  (\cos(\theta) + \sin(\theta)) \sin(\varphi) $
peaks in the $yz$ plane at an angle $\theta = \pi/4$.
It would be subject to a nonzero torque in the $x$ direction.

The torque on the orientation of a fission fragment does not affect
the angular momentum in the same way.  The expectation value of 
the torque vanishes because 
the wave function's probability distribution
is independent of the azimuthal angle $\varphi$.
To understand this, note that $|\Psi(\theta,\varphi)| =
|\Psi(\theta,\varphi+\pi)|$, so the contributions at both points will cancel.
However, the Coulomb field still manages to increase the angular momentum 
content even when the probability distribution is axially symmetric.
The change is not visible in  $\langle \hat L_x \rangle $ or $\langle \hat L_y\rangle$ 
but rather through the squared total angular momentum in 
the {xy} plane, $   \hat L_x^2 + \hat L_y^2 $.
For convenience, we assume a wave packet independent of $\varphi$, 
 and  we cast
the polar angular variable from  $\theta$ to $\mu = \cos(\theta)$,
\begin{align}
\hat L_x^2 + \hat L_y^2 = -\frac{\partial}{\partial \mu} (1-\mu^2)
\frac{\partial}{\partial \mu}.
\end{align}
Its  commutator with the angular part of the Coulomb field is given by
\begin{align}
[\hat L_x^2 + \hat L_y^2, P_2(\mu) ] =
-3 + 9\mu^2 + 6(\mu^3 - \mu)\frac{\partial}{\partial \mu}. 
\end{align}
The commutator is a real anti-Hermitian operator so its diagonal
matrix elements in real-valued wave functions are zero.  Thus there
is no change in $L^2$ for the initial wave function as we have constructed.
The wave function becomes complex during the time evolution due to the
kinetic operator that changes the relative phases of the $L$ components.
This links the change in angular momentum to the
moment of inertia.  Smaller moments of inertia
allow the relative phases of the components to build up more quickly,
allowing the early transient Coulomb field to play a stronger role.
 
In the reaction producing $^{108}$Ru ($\beta_2~=~0.82$), the Coulomb excitation 
plays an important role due to the large deformation of the fragment.
As a result, the average angular momentum increases from 
 $9.3~\hbar$ at the scission to $12.3~\hbar$ in the final state. This result shows the importance of the fission dynamics since a part of the angular momentum is generated in the Coulomb reorientation phase. 
 
\section{The angle of the fission axis}
\label{sec:ang_fis_axis}

In general, the wave function of the collective model depends on an additional
angle, namely the solid angle $(\theta_f, \varphi_f)$  of the fission axis 
in the overall center-of-mass 
frame (CMF).  We assume in our Hamiltonians that this angle is redundant
provided that the overall angular momentum of the fissioning nucleus is
zero.  In this Appendix, we outline how this can be proved for calculating
$V_{NN}$ in the single-angle model.  

First of all, the opening angle $\theta_{ab}$ between two
axes in terms of their angles in the CMF frame is given by
\begin{equation}
\theta_{ab} = {\rm arccos} \left( \cos \theta_a \cos \theta_b +
\sin \theta_a \sin \theta_b \cos (\varphi_a -\varphi_b)\right)
\end{equation}
where $(\theta_i,\varphi_i)$ are the spherical coordinates 
of an axis $i$ in the CMF.
Then, with the help of the identity \cite{1}
we can write the 
spherical harmonic $Y_{L0}(\theta_{ab})$ as
\begin{equation}
Y_{L0}(\theta_{ab}) = \left(\frac{4 \pi}{2 L +1}\right)^{1/2}\sum_M
Y_{LM}(\theta_a,\varphi_a)Y^*_{LM}(\theta_b,\varphi_b)
\end{equation}
$$
=(4 \pi)^{1/2} [L_a L_b]^0
$$
where $[L_1,L_2]^{L_3}$ is the
usual notation for coupled angular momenta.
The matrix element $\langle L_1 | Y_{L_{ab}}| L_2\rangle$ in
the single-angle representation can then be expressed in the
CMF by using the above identity for each of the spherical harmonics.  After
carrying out some angular momentum algebra \cite{2}, the integral
over the fission axis separates out and can be evaluated
analytically.  The result is identical to the evaluated
matrix element in the single-angle representation.

\end{document}